\documentclass[twocolumn,preprintnumbers,amsmath,amssymb,aps,prb]{revtex4}
\usepackage{graphicx}
\begin{document}

\title{
Clogging and Jamming Transitions in Periodic Obstacle Arrays 
} 
\author{
H.T. Nguyen$^{1,2}$, C. Reichhardt$^{1}$, and  C. J. Olson Reichhardt$^{1}$}
\affiliation{
$^{1}$Theoretical Division,
  Los Alamos National Laboratory, Los Alamos, New Mexico 87545 USA\\
  $^{2}$Department of Physics,
  University of South Florida, Tampa, Florida 33620 USA
} 

\date{\today}
\begin{abstract}

We numerically examine clogging transitions for bidisperse disks
flowing through a two dimensional periodic obstacle array.
We show that clogging is a probabilistic event that occurs
through a transition from a homogeneous
flowing state to a heterogeneous or phase separated jammed state where
the disks form dense connected clusters.
The probability for clogging to occur during a fixed time
increases with increasing particle packing  and obstacle number.
For driving at different angles with respect to the symmetry
direction of the obstacle array,
we show that certain directions have a higher clogging susceptibility.
It is also possible to have a size-specific clogging transition
in which one disk size becomes completely immobile
while the other disk size continues to flow.
\end{abstract}
\maketitle

\vskip2pc
A loose collection of
particles such as grains or bubbles can exhibit a transition
from a flowing liquidlike state to
a non-flowing or jammed state as a function of increasing density, where
the density $\phi_{j}$ at which the system jams is referred to as Point J \cite{1,2,3}.
One system in which jamming has been extensively studied is
a bidisperse two-dimensional (2D) packing of frictionless disks,
where the area faction covered by the disks at Point J is
approximately $\phi=0.84$, and where the system density is 
uniform at the jamming transition \cite{1,2,4}.
Related to jamming is the phenomenon of clogging, as observed in 
the flow of grains \cite{5,6,7,8} or bubbles \cite{9}
through an aperture at the tip of a  hopper.
The clogging transition is a
probabilistic process in which, for a  fixed grain
size, the probability of a clogging event occurring during a fixed time interval
increases with decreasing aperture size. 
A general question is whether there are systems that can exhibit
features of both jamming and clogging.
For example, such combined effects could appear
in a system containing quenched disorder such as pinning or obstacles
where jammed or clogged configurations
can be created by a combination of particles that
are directly immobilized in a pinning site 
as well as other particles that are indirectly immobilized through contact with
obstacles or pinned particles.
In many systems where pinning effects arise,
such as for superconducting vortices
or charged particles, the
particle-particle interactions are long range, meaning that there is no
well defined areal coverage density \cite{10} at which the system can be said
to jam, so a more ideal system to study is an assembly of hard disks with strictly
short range particle-particle interactions.
Previous studies have considered
the effect of a random pinning landscape on transport in a
2D sample of bidisperse hard disks \cite{11},
while in other work on the effect of obstacles,
the density at which jamming occurs decreases when the
number of pinning sites or obstacles increases \cite{12,13}.

Here we examine a 2D system of bidisperse frictionless disks
flowing through a square periodic
obstacle array composed of immobile disks
with an obstacle lattice constant of $a$.
The
total disk density, defined as the
area coverage of the mobile disks and the obstacles, is $\phi_{t}$.  We find that
for $\phi_t$ far below the obstacle-free jamming density $\phi_{j}$, the system can
reach clogged configurations by forming a phase separated state
consisting of a high density
connected cluster surrounded by
empty regions,
and that the clogging probability $P_c$ during a fixed time
interval depends on both $a$ and $\phi_{t}$.
There is also a strong dependence of $P_c$ 
on the direction of drive with respect to the
obstacle lattice symmetry,
with an increase in $P_c$ for 
certain incommensurate angles.
At finite drive angles we
observe a novel size-dependent clogging
effect in which the smaller disks become completely jammed while
a portion of the larger disks continue to flow.
This work is relevant for filtration processes \cite{14,15,16},
the flow of discrete particles in porous media \cite{17,18}, and the flow and
separation of of colloids
on periodic substrates \cite{19,20,21,22}

{\it Model and Method---}
We consider a 2D square system of size $L \times L$
where $L = 60$ with periodic boundary conditions in the $x$ and $y$-directions.
The sample contains $N_l$ disks of diameter $\sigma_{l} = 0.7$ and
$N_s=N_l$ disks of diameter $\sigma_{s} = 0.5$, giving a size ratio of
$1:1.4$.  This same size ratio has been studied in previous works
examining jamming in bidisperse obstacle-free disk packings, where
jamming occurs near a packing fraction
of $\phi_j = 0.844$
and is associated with a contact number of $Z = 4.0$ \cite{1,2,3,4}.
A total of $N_p$  obstacles are placed in a square lattice with lattice constant $a$ and are 
modeled as disks of diameter $\sigma_s=0.5$ held at fixed positions.
The initial configuration is prepared by
placing the small and large disks in non-overlapping random positions with a uniform density.
The disks interact through a repulsive short range harmonic force,
${\bf F}^{in}_{ij} =k(\sigma_{ij} - |{\bf r}_{ij}|)\Theta(\sigma_{ij} - |{\bf r}_{ij}|){\hat {\bf r}}_{ij}$
where $\sigma_{ij}=(\sigma_i+\sigma_j)/2$ is the sum of the radii of disks $i$ and $j$,
${\bf r}_{ij}={\bf r}_i-{\bf r}_j$, ${\hat {\bf r}}_{ij}={\bf r}_{ij}/|{\bf r}_{ij}|$,
$\Theta$ is the Heaviside step function, and
the interaction force drops to zero when the
separation $|{\bf r}_{ij}|>\sigma_{ij}$.
The spring constant is set to $k=300$ which
is large enough to ensure that the overlap between disks for the largest
driving force considered in this work remains small.
After initialization we apply a constant  driving force
${\bf F}_{d}$ to the mobile disks
which could
arise from a gravity or fluid induced flow.
The dynamics for a given disk $i$ at
position ${\bf r}_{i}$ is obtained by integrating the following
overdamped equation of motion:
\begin{equation}
\eta \frac{d {\bf r}_{i}}{dt} = \sum^{N}_{i\ne j}{\bf F}^{in}_{ij} + {\bf F}_d  \ .
\end{equation}
Here $N=N_s+N_l+N_p$ is
the total number of disks and the damping constant $\eta$ is set to unity.
The external driving force is given by
${\bf F}_{d} = F_d(\cos(\theta){\bf \hat x} + \sin(\theta){\bf \hat y})$,
where  $\theta$ is the angle of the driving direction with respect to the positive
$x$ axis.  We take $F_d=0.025$ but, provided $F_d$ is sufficiently small, our results
are not sensitive to the choice of $F_d$.
In the absence of obstacles, all the disks move in the driving direction at a speed of
$F_d$.
The total disk density $\phi_t$ is the area fraction covered by the free disks and obstacles,
$\phi_{t} = \frac{1}{4} \pi (N_l \sigma^2_{l} + (N_s+N_p)\sigma^2_{s})/L^2$.
To quantify the clogging transition, we monitor the average velocity of  the mobile
disks along the $x$ and $y$ directions,
$\langle V_{x,y} \rangle  =( N_s+N_l)^{-1}\sum_{i = 1}^{N_s+N_l} {\bf v}_{i}\cdot (\hat{\bf x}, \hat{\bf y})$,
where ${\bf v}_i$ is the velocity of disk $i$.
To ensure that the system has reached a steady state,
we run all simulations for $3 \times 10^8$
simulation time steps and average the values of
$\langle V_x \rangle$ and $\langle V_y \rangle$ over $10^5$ simulation time steps.
We define $P_c$ to be the probability that the system will
reach a clogged state                                               
with  $\langle V_{x}\rangle = 0.0$ after
a total of $3\times 10^8$ simulation time steps,
and perform
100 realizations 
for each value of $\phi_t$ and $a$.

\begin{figure}
\includegraphics[width=3.5in]{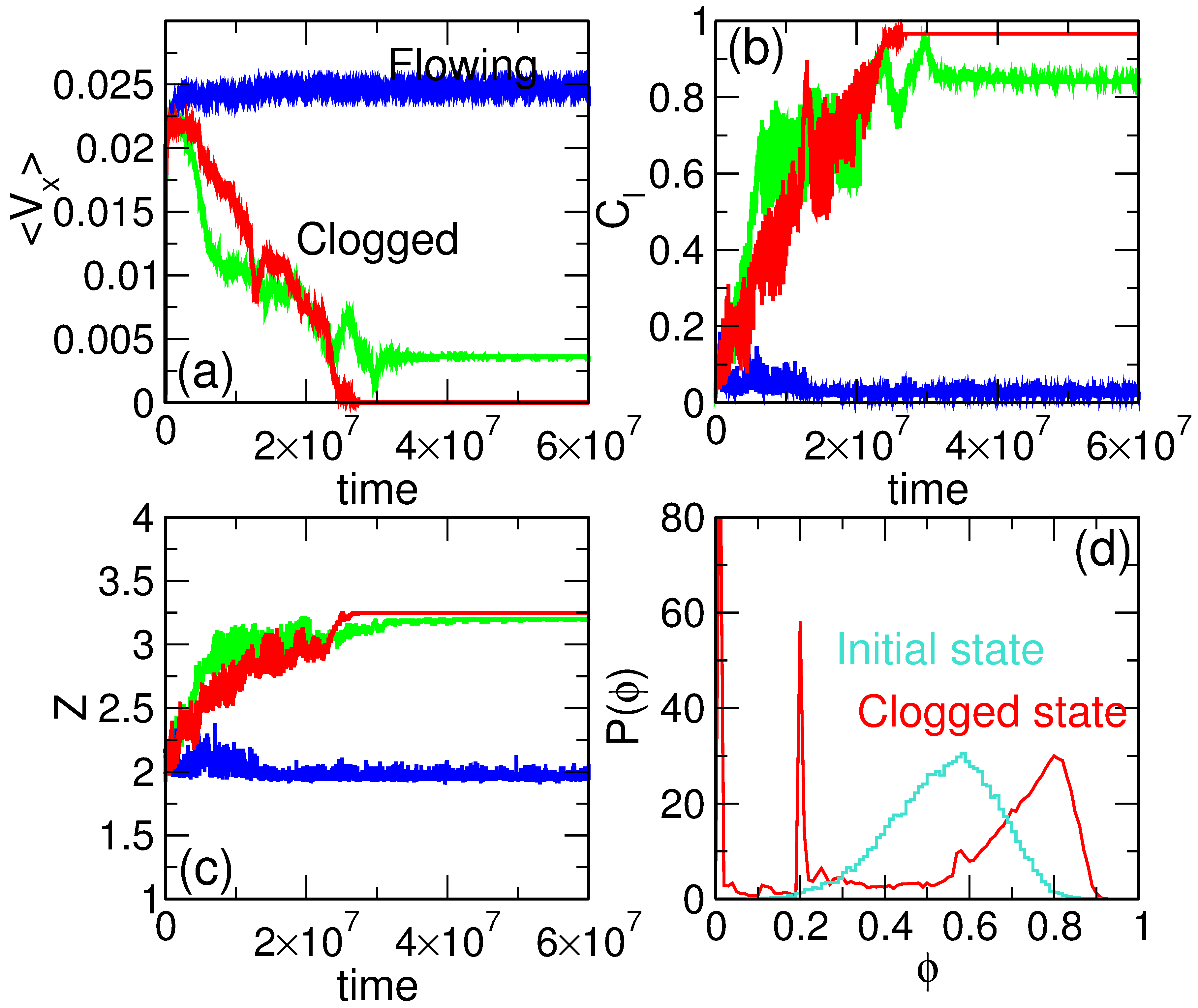}
\caption{ (a) The average disk velocities $\langle V_{x}\rangle$,
  (b) the fraction of disks in a cluster  $C_{l}$, and (c)
  the average contact number $Z$ versus time in simulation time steps
  for a 2D system of bidisperse  disks moving through a square
  periodic obstacle array with total disk density
  of $\phi_t=0.54$, lattice constant of $a = 3.0$, and
  constant external drive $F_d=0.025$ applied in the
  positive $x$-direction. Blue curves: a run in which the disks remain flowing;
  red curves: a run in which the disks become clogged;
  green curves: a run in which the disks become partially clogged.
  (d) Distribution $P(\phi)$ of local disk density $\phi$ in the initial state (blue) and
  after reaching a clogged state (red), averaged over 40 clogged realizations.
}
\label{fig:1}
\end{figure}

{\it Results---}
We fist consider the case in which the external drive is applied along the
$x$ direction with $\theta=0$.
In Fig.~\ref{fig:1}(a) we plot $\langle V_{x}\rangle$ versus time for a system
with $\phi_{t} = 0.54$ and $a = 3.0$.
At this obstacle density, we find that the clogging probability
$P_c = 1.0$ for  $\phi_{t} > 0.62$, $P_c \approx 0$
for $\phi_t< 0.52$, 
and at $\phi_t=0.54$, the density shown in the figure,
$P_c = 0.31$.
We illustrate three representative realizations in Fig.~\ref{fig:1}(a,b,c): one in which
the system does not clog but continues to flow, one in which the system clogs completely,
and one in which a partial clogging occurs where at least three-quarters of the disks
are no longer moving.
Due to
the nonequilibrium fluctuations,
it is possible that if we were to consider a longer time average,
the flowing or partially clogged states may fully clog; however, the fully
clogged states can never unclog.
In realizations that reach a clogged state, the
system does not pass instantly from a
flowing to a non-flowing state, but instead exhibits a series of steps
in which a progressively larger number of disks become clogged, with
$\langle V_x\rangle$ continuing to diminish until it reaches zero.
This behavior is different from that typically observed in hopper flows, where   
a single event brings the flow to a sudden and complete halt.
The red curve in Fig.~1(a)
contains time intervals during which
the number of flowing grains, which is
directly proportional to the value of $\langle V_x\rangle$,
temporarily increases prior to the system reaching a final clogged
state with $\langle V_x\rangle=0$ after
$2.5 \times 10^7$ simulation time steps.
Since there are no thermal fluctuations or  external vibrations,
once the system is completely clogged, all of the dynamical fluctuations
disappear and the system
is permanently  absorbed into  a clogged state.            
In Fig.~\ref{fig:1}(b) we plot
the fraction $C_l$ of mobile disks that are in
the largest connected cluster versus time,
while in
Fig.~\ref{fig:1}(c) we show the corresponding average disk contact number $Z$.
For the realization that fully clogs, 
$C_{l}$ gradually increases with time, indicating
that there is a single growing cluster, while $Z$ also increases.
When $\langle V_{x}\rangle$ reaches zero,
$C_{l} = 0.98$, indicating that almost all the disks have formed a
single cluster, while $Z = 3.25$,
which is well below the critical value $Z_{c} = 4.0$ expected at the obstacle-free
jamming transition.
In contrast, for the system
that remains flowing, $\langle V_{x}\rangle = 0.025$,
indicating that almost all of the mobile grains are
freely flowing.  At the same time, $C_{l}$ is close to zero and $Z = 2.0$.

\begin{figure}
\includegraphics[width=3.5in]{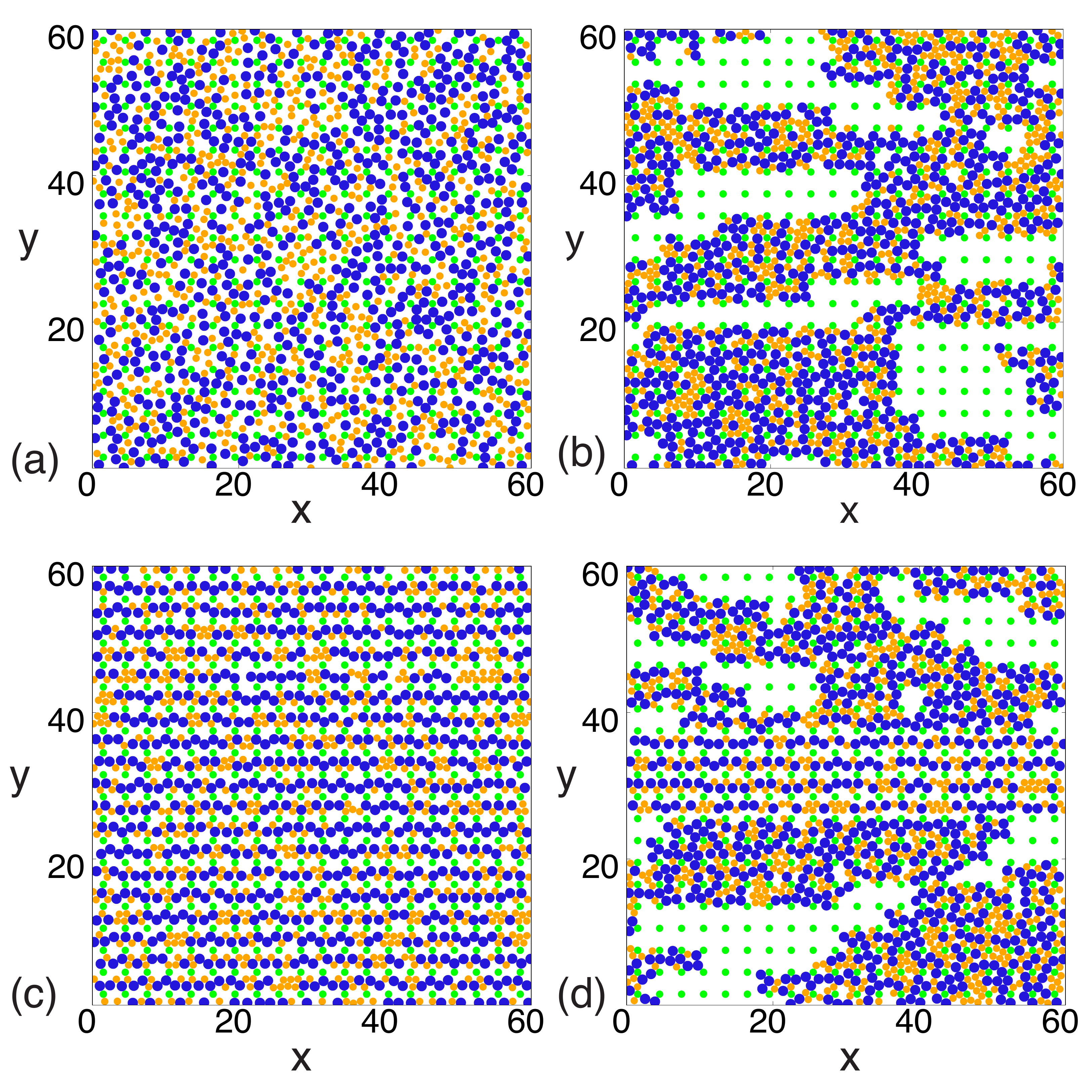}
\caption{ Images of the obstacle locations
  (green circles) and the mobile disks (large disks: blue; small disks: orange)
  for the samples shown in Fig.~\ref{fig:1} from systems
  with
  $a=3.0$
  and $\phi_{t} = 0.54$. (a) Initial configuration of the sample that clogs.
  (b) Final clogged configuration of the same sample.
  (c) Late time snapshot of the sample that continues to flow.
  (d) Late time snapshot of the sample that partially clogs.
}
\label{fig:2}
\end{figure}

In Fig.~\ref{fig:2}(a)
we show an image of the initial uniform density disk configuration
for the system in Fig.~\ref{fig:1}(a) which reaches a clogged state.
For the same sample, Fig.~\ref{fig:2}(b) illustrates the clogged state
with $\langle V_{x}\rangle = 0$.  The disks phase separate into
a high density connected cluster surrounded by
regions devoid of mobile disks.
In contrast, Fig.~\ref{fig:2}(c) shows a late time image of the sample from Fig.~\ref{fig:1}(a)
that remains flowing.  Here the overall disk density is uniform
and the motion is confined in one-dimensional (1D) channels that run between the
rows of obstacles.  For the partially clogged state, Fig.~\ref{fig:1}(b)
indicates that the cluster fraction
$C_l=0.84$ is lower than the value $C_l=0.98$ observed in the fully
clogged state.  At late times for the partially clogged sample, Fig.~\ref{fig:2}(d) shows that 
a large jammed cluster forms,
while in the middle of the sample there is a region of uniform disk density
through which the grains flow in 1D channels.
Thus, the partially clogged state combines features of the
clogged and flowing states in Fig.~\ref{fig:2}(b,c).

In Fig.~\ref{fig:1}(d) we plot the distribution $P(\phi)$ of the local packing density
$\phi$ at initial and late times for a sample that reaches a clogged state.  To measure
$\phi$, we divide the sample into squares of size $2 \times 2$ and find the area fraction
of each square covered by free disks and obstacles.  In the initial state, there is a peak
in $P(\phi)$ centered at the total disk density of $\phi_t=0.54$.  In contrast,
$P(\phi)$ has multiple peaks in the clogged state centered at $\phi=0$, corresponding to
empty regions, $\phi=0.2$, corresponding to the obstacle density, and
at $\phi=0.82$, corresponding to the clogged regions which have a density close to
the free disk jamming density.

\begin{figure}
\includegraphics[width=3.5in]{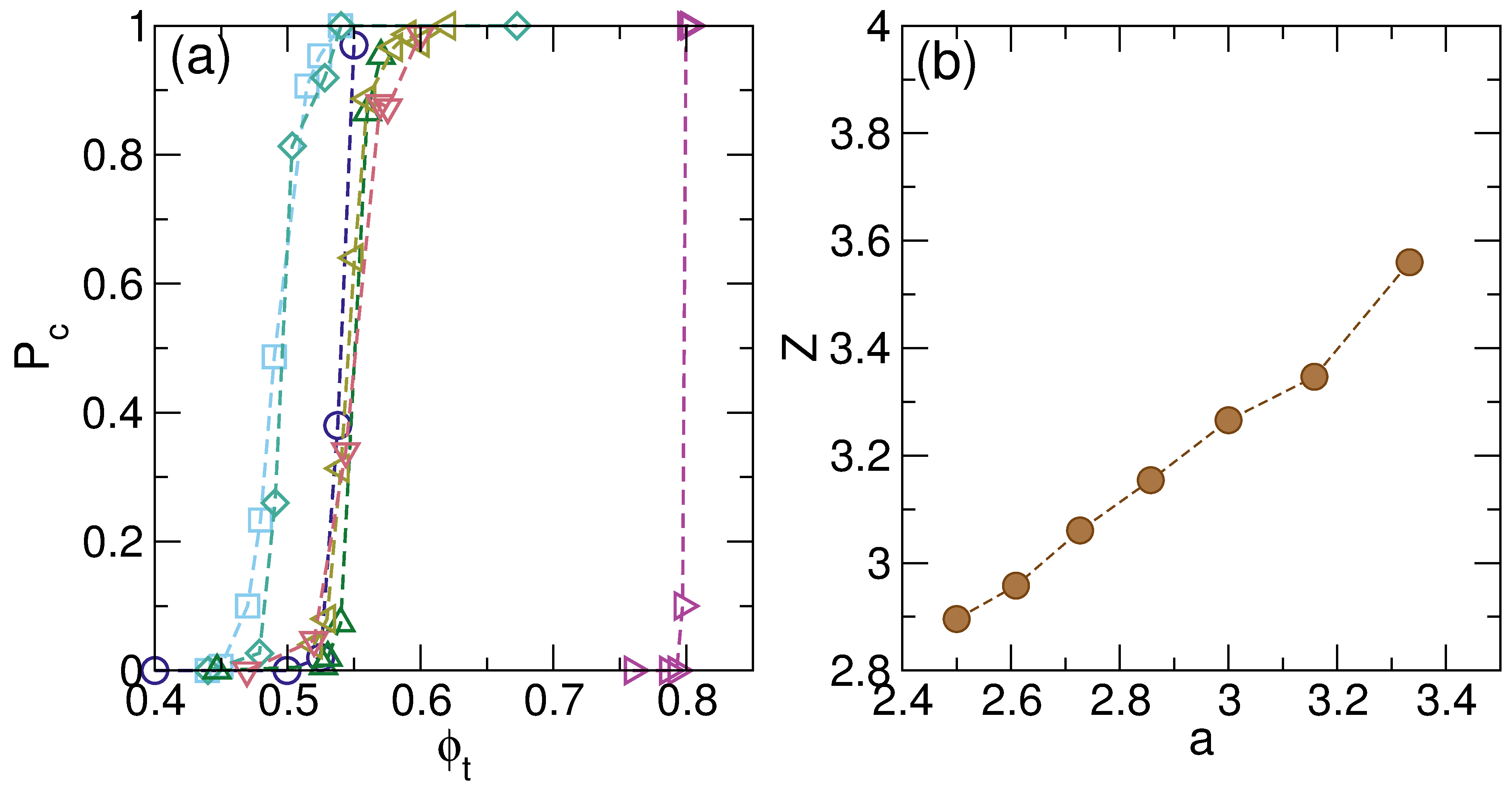}
\caption{(a) The fraction $P_c$ of states that clog vs
  $\phi_{t}$ for varied obstacle lattice constant $a=2.5$ (dark blue circles),
  2.609 (light blue squares), 2.727 (light green diamonds), 2.857 (dark green up triangles),
  3.0 (orange left triangles), 3.158 (red down triangles), and 3.333
  (magenta right triangles).
(b) The average value of $Z$ for realizations that clog vs $a$ showing a linear
increase in $Z$ with $a$.
}
\label{fig:3}
\end{figure}

In Fig.~\ref{fig:3}(a) we plot the clogging probability $P_c$ 
versus $\phi_{t}$
for samples with obstacle lattice constant ranging from $a = 2.5$
to $a=3.33$.  We perform
100 realizations for each value of $\phi_{t}$.
When $a = 3.33$, $P_c = 0$
for $ \phi_{t} < 0.79$, and there is a sharp
increase to $P_c =1.0$ at $\phi_{t} = 0.8$, indicating that when the
spacing between obstacles is large,
a high density of mobile particles must be introduced in order for the
system to clog.
We take the critical density $\phi_t^c$ to be the value of $\phi_t$ at which
$P_c$ passes through $P_c=0.5$.
As $a$ decreases, $\phi_t^c$ also decreases, and at $a=2.5$, $\phi_t^c=0.49$.
We do not observe a strictly monotonic decrease in 
$\phi^{c}_{t}$ with decreasing $a$ since for some
values of $a$ there are particular
combinations of disk configurations that can better fit in
the constraint of a square obstacle lattice.
Since our sample size is
fixed at a finite value, the square symmetry of our obstacle lattice constrains
us to a discrete number of possible choices of $a$.
By averaging the contact number $Z$ over only realizations that clog, we find
a monotonic increase $Z$ with $a$, as shown 
in Fig.~\ref{fig:3}(b),
where $Z$ increases from $Z=2.9$ at $a = 2.5$ to $Z=3.6$ at $a = 3.33$.             
In principle, $Z$ will approach the value $Z=4.0$ for very large values of $a$ or in the
limit of a single obstacle when 
$\phi_t=\phi_{j} \approx 0.84$;
however, the time required to reach clogged states at large $a$ increases well beyond
the length of our fixed simulation time window.

\begin{figure}
\includegraphics[width=3.5in]{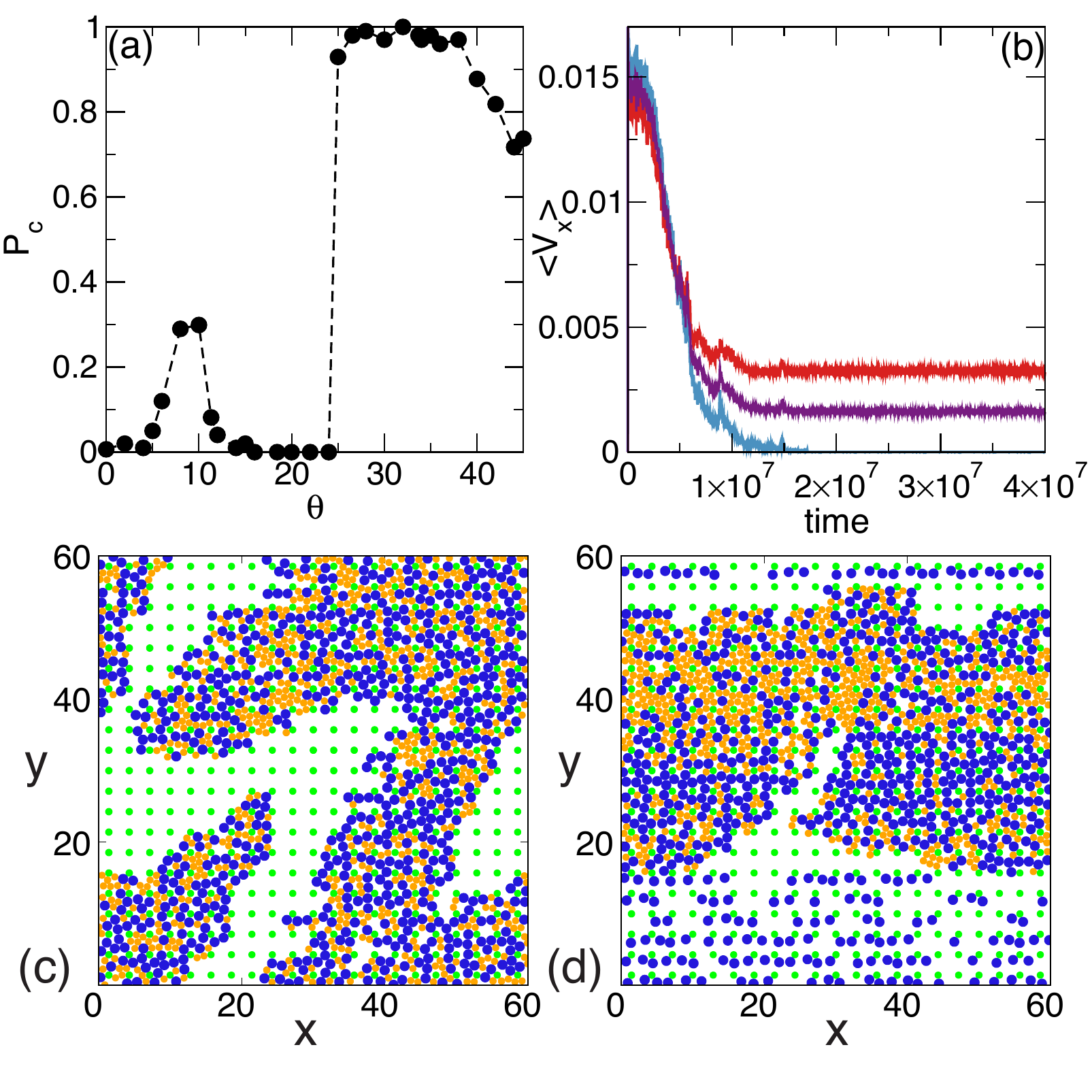}
\caption{(a) The fraction $P_c$ of states that clog vs $\theta$, the angle the driving
  direction makes with the positive $x$ axis, in samples with
  $\phi_t = 0.5272$ and $a = 2.857$.
  The susceptibility to clogging is enhanced
  for $\theta > 25^{\circ}$.
  (b) $\langle V_x\rangle$ vs time in simulation time steps for the
  large disks only (red), the small disks only (blue), and all disks (purple) for a driving
  angle of
  $\theta = 20^{\circ}$.
  We find a size dependence, with only the smaller disks becoming clogged while the
  large disks continue to flow.
  (c) The disk configuration in the clogged state at $\theta = 32^{0}$ from
  panel (a). (d) The disk configuration
for the size-dependent clogged state
from panel (b).
}
\label{fig:4}
\end{figure}

{\it Directional dependence and size dependent clogging---}
We next consider the effect of changing the direction of the drive relative to the
symmetry axes of the square obstacle array.
In Fig.~\ref{fig:4}(a) we plot
$P_c$
versus the drive angle $\theta$
in samples with $\phi_t = 0.527$ and $a = 2.857$.
For each value of $\theta$, we perform
100 realizations.
Here,
$P_c = 0$ for $\theta = 0$, as also shown in
Fig.~\ref{fig:3}(a).
As $\theta$ increases, a local maximum in $P_c$ with $P_c=0.3$ appears near
$\theta = 10$.
This is followed by a drop to $P_c=0$ over the range
$15^{\circ} < \theta < 25^{\circ}$, and an increase to
$P_c=0.98$ for $25^{\circ} \leq \theta < 40^{\circ}$,
with a dip to $P_c=0.72$ occurring near $\theta  = 45^{\circ}$.
Due the symmetry of the obstacle lattice,
the same features repeat over the range $45^{\circ}<\theta < 90^{\circ}$.
The increase of $P_c$ near $\theta  = 10^{\circ}$
occurs due to a break down of the 1D channeling that arises
for the $\theta = 0^{\circ}$ flow.
Similarly, the dip in $P_c$ near $\theta=45^{\circ}$
appears when the disks become able to form 1D channels of flow along the
diagonal direction.
As $\theta$ varies, we find that
certain angles, such as $\theta=0^{\circ}$ and $\theta=45^{\circ}$,
allow 1D channeling motion,
whereas at drive angles of $25^{\circ}<\theta<40^{\circ}$ there is no easy flow direction
so the disks are forced to collide with the obstacles,
producing an increase in $P_c$.
In Fig.~\ref{fig:4}(c) we illustrate a clogged state
that is aligned with the driving angle of $\theta = 32^\circ$.

For
$20^{\circ} \leq \theta \leq 24^{\circ}$ we observe a
size-dependent clogging behavior in which
the smaller disks become completely clogged while
a portion of the larger
disks continue to flow.
In Fig.~\ref{fig:4}(b) we plot $\langle V_x\rangle$ for the large and small disks separately
and for all disks combined for a driving angle of
$\theta = 20^{\circ}.$
After $2\times 10^7$ simulation time steps,
for the small disks $\langle V_{x}\rangle = 0$,
indicating the complete clogging of the small
disks along with a saturation to a steady state flow
for the larger disks.
This result is counter-intuitive since it might be expected
that the larger disks would clog first.
In Fig.~\ref{fig:4}(d) we show a snapshot of the size-dependent clogged state
from Fig.~\ref{fig:4}(b).
All of the smaller disks
are jammed in a  cluster along with a portion of the larger disks,
while in the lower density regions there are a number
of larger disks undergoing channeling motion along the $x$-direction.
 The size-dependent clogging can be understood as a consequence of a
 directional locking effect \cite{19,20,21,22,23}
 in which the flow of the larger disks
 remains locked to the $\theta = 0^{\circ}$ direction of the
 obstacle lattice
 while the flow of the smaller disks follows the angle of the drive,
 which increases the chance for the smaller disks to become clogged.
 For $\theta$ just below $\theta=20^\circ$, most of the smaller disks
 are clogged but there are a few that remain mobile.
 The directional locking effect,
 in which particles preferentially move along
 lattice symmetry directions,
 has been observed for colloids \cite{19,20,21,22} and
 superconducting vortices \cite{23} moving over periodic
 substrates.
 It can be used to perform particle separation
 by having one species lock to a symmetry direction
 while the other does not.
 In our case, the disk size that does not undergo directional locking
 ends up in a clogged state,
 suggesting that species separation by selective clogging
 could be a new method for
particle separation.

{\it Conclusion---} 
We have investigated the clogging transition for a bidisperse assembly of
frictionless disks moving through a two-dimensional square obstacle array.
We find that the probability of clogging during a
fixed time interval
increases with increasing total disk density $\phi_t$ and
decreases with the obstacle spacing $a$.
For disk densities well below the obstacle-free jamming density,
the clogged states are phase separated and consist of
a connected high density jammed cluster surrounded by
a low density disk-free region.
In the clogged state, the contact number $Z$ increases
linearly with decreasing obstacle density.
We also find that the
clogging  probability has a strong dependence on the relative angle
between the driving direction and the symmetry axes of the square obstacle array.
The clogging is enhanced for incommensurate
angles such as $\theta=35^{\circ}$ where the 1D channeling
flow of the disks between the obstacles is suppressed.
We also find that for some drive angles
there is a size-dependent clogging effect
in which the smaller disks become completely clogged while a portion of the larger
disks remain mobile.
Here the motion of the larger disks remains locked
along the $x$-axis of the obstacle array
whereas the smaller disks move in the driving direction.
This suggests that selective clogging could be used as a particle separation method. 
\acknowledgments
This work was carried out under the auspices of the 
NNSA of the 
U.S. DoE
at 
LANL
under Contract No.
DE-AC52-06NA25396.
H.N. gratefully acknowledges support from NSF Grant No. DMR-1555242.


\begin{thebibliography}{99}

\bibitem{1}
  C.S. O'Hern, L.E. Silbert, A.J. Liu, and S.R. Nagel,
Jamming at zero temperature and zero applied stress: The epitome of disorder,
Phys. Rev. E {\bf 68}, 011306 (2003).

\bibitem{2}
  J.A. Drocco, M.B. Hastings, C.J. Olson Reichhardt, and C. Reichhardt,
Multiscaling at Point J: Jamming is a critical phenomenon,
Phys. Rev. Lett. {\bf 95}, 088001 (2005).

\bibitem{3}
  A.J. Liu and S.R. Nagel,
The jamming transition and the marginally jammed solid,
Annu. Rev. Condens. Matter Phys. {\bf 1}, 347 (2010).

\bibitem{4}
C. Reichhardt and C. J. Olson Reichhardt,
Aspects of jamming in two-dimensional athermal frictionless systems,
Soft Matter {\bf 10}, 2932 (2014).

\bibitem{5}
  K. To, P.-Y. Lai, and H.K. Pak,
Jamming of granular flow in a two-dimensional hopper,
Phys. Rev. Lett. {\bf 86}, 71 (2001).

\bibitem{6}
C.C. Thomas and D.J. Durian,
Geometry dependence of the clogging transition in tilted hoppers,
Phys. Rev. E {\bf 87}, 052201 (2013).

\bibitem{7}
  A. Garcimart{\' \i}n, J.M. Pastor, L.M. Ferrer, J.J. Ramos, C. Mart{\' \i}n-G{\' o}mez,
  and I. Zuriguel,
Flow and clogging of a sheep herd passing through a bottleneck,
Phys. Rev. E {\bf 91}, 022808 (2015).

\bibitem{8}
  I. Zuriguel, L.A. Pugnaloni, A. Garcimart{\' \i}n, and D. Maza,
Jamming during the discharge of grains from a silo described as a percolating transition,
Phys. Rev. E {\bf 68}, 030301(R) (2003).

\bibitem{9}
  D. Chen, K.W. Desmond, and E.R. Weeks,
  Topological rearrangements and stress fluctuations in quasi-two-dimensional hopper
  flow of emulsions,
Soft Matter {\bf 8}, 10486 (2012).

\bibitem{10}
 C. Reichhardt and C.J. Olson Reichhardt,
 Depinning and nonequilibrium dynamic phases of particle assemblies driven over random
 and ordered substrates: a review,
 Rep. Prog. Phys., in press.

\bibitem{11}
C.J. Olson Reichhardt, E. Groopman, Z. Nussinov, and C. Reichhardt,
Jamming in systems with quenched disorder,
Phys. Rev. E {\bf 86}, 061301 (2012).

\bibitem{12}
C. Brito, G. Parisi, and F. Zamponi,
Jamming transition of randomly pinned systems,
Soft Matter {\bf 9}, 8540 (2013).

\bibitem{13}
  A.L. Graves, S. Nashed, E. Padgett, C.P. Goodrich, A.J. Liu, and J.P. Sethna,
Pinning susceptibility: The effect of dilute, quenched disorder on jamming,
Phys. Rev. Lett. {\bf 116}, 235501 (2016).

\bibitem{14}
S. Redner and S. Datta,
Clogging time of a filter,
Phys. Rev. Lett. {\bf 84}, 6018 (2000).

\bibitem{15}
F. Chevoir, F. Gaulard, and N. Roussel,
Flow and jamming of granular mixtures through obstacles,
EPL {\bf 79}, 14001 (2007).

\bibitem{16}
N. Roussel, T.L.H. Nguyen, and P. Coussot,
General probabilistic approach to the filtration process,
Phys. Rev. Lett. {\bf 98}, 114502 (2007).

\bibitem{17}
D. Liu, P.R. Johnson, and M. Elimelech,
Colloid deposition dynamics in flow through porous media: Role of electrolyte concentration,
Environ. Sci. Technol. {\bf 29}, 2963 (1995).

\bibitem{18}
F. Wirner, C. Scholz, and C. Bechinger,
Geometrical interpretation of long-time tails of first-passage time distributions in porous
media with stagnant parts,
Phys. Rev. E {\bf 90}, 013025 (2014).

\bibitem{19}
  P.T. Korda, M.B. Taylor, and D.G. Grier,
Kinetically locked-in colloidal transport in an array of optical tweezers,
Phys. Rev. Lett. {\bf 89}, 128301 (2002).

\bibitem{20}
M.P. MacDonald, G.C. Spalding, and K. Dholakia,
Microfluidic sorting in an optical lattice,
Nature (London) {\bf 426}, 421 (2003).

\bibitem{21}
Z. Li and G. Drazer,
Separation of suspended particles by arrays of obstacles in microfluidic devices,
Phys. Rev. Lett. {\bf 98}, 050602 (2007).

\bibitem{22}
A.M. Lacasta, J.M. Sancho, A.H. Romero, and K. Lindenberg,
Sorting on periodic surfaces,
Phys. Rev. Lett. {\bf 94}, 160601 (2005).

\bibitem{23}
C. Reichhardt and F. Nori,
Phase locking, devil's staircases, Farey trees, and Arnold tongues in driven vortex
lattices with periodic pinning,
Phys. Rev. Lett. {\bf 82}, 414 (1999).

\end{thebibliography}
\end{document}